\documentclass[12pt]{iopart}
\usepackage{graphicx}
\usepackage{soul}
\graphicspath{{figures/}}

\def\ltsima{$\; \buildrel < \over \sim \;$}
\def\ltsim{\lower.5ex\hbox{\ltsima}}
\def\gtsima{$\; \buildrel > \over \sim \;$}
\def\gtsim{\lower.5ex\hbox{\gtsima}}
\def\simlt{\mathrel{\hbox{\rlap{\hbox{\lower4pt\hbox{$\sim$}}}\hbox{$<$}}}}
\def\simgt{\mathrel{\hbox{\rlap{\hbox{\lower4pt\hbox{$\sim$}}}\hbox{$>$}}}}

\def\nr#1{numerical relativity#1 (NR#1)\gdef\nr{NR}}

\def\movingpuncture{\emph{moving puncture}\gdef\movingpuncture{moving puncture}}

\def\newacronym#1#2#3{\gdef#1{#3 (#2)\gdef#1{#2}}}

\newacronym{\xirel}{XiRel}{Next Generation Infrastructure for Numerical Relativity}
\newacronym{\cigr}{CIGR}{Community Infrastructure for General Relativistic
MHD}
\newacronym{\mpm}{MPM}{moving puncture method}
\newacronym{\ghm}{GHM}{generalized harmonic method}
\newacronym{\sph}{SPH}{smooth particle hydrodynamics}
\newacronym{\amr}{AMR}{adaptive mesh refinements}
\newacronym{\NSF}{NSF}{National Science Foundation}
\newacronym{\NASA}{NASA}{National Aeronautics and Space
Administration}
\newacronym{\lisa}{LISA}{Laser Interferometer Space Antenna}
\newacronym{\ligo}{LIGO}{Laser Interferometer Gravitational-wave
  Observatory}
\newacronym{\cra}{CRA}{Center for Relativistic Astrophysics}
\newacronym{\Caltech}{Caltech}{California Institute of Technology}
\newacronym{\MIT}{MIT}{Massachusetts Institute of Technology}
\newacronym{\sph}{SPH}{smooth particle hydrodynamics}
\newacronym{\tsi}{TSI}{the Terascale Supernova Initiative}
\newacronym{\wmap}{WMAP}{the Wilkinson Microwave Anisotropy Probe}
\newacronym{\decigo}{DECIGO}{the Deci-Hertz Interferometric
  Gravitational-wave Observatory}
\newacronym{\cmbr}{CMBR}{cosmic microwave background}
\newacronym{\ibbh}{IBBH}{intermediate binary black hole}
\newacronym{\bdj}{BDJ}{Brans-Dicke-Jordan}
\newacronym{\bbo}{BBO}{Big Bang Observer}
\newacronym{\decigo}{DECIGO}{Deci-Hertz Gravitational-Wave
  Observatory}
\newacronym{\gr}{GR}{general relativity}
\newacronym{\grhd}{GRHD}{general relativistic hydrodynamics}
\newacronym{\grmhd}{GRMHD}{general relativistic magnetohydrodynamics}
\newacronym{\gw}{GW}{gravitational wave}
\newacronym{\EM}{EM}{electromagnetic}
\newacronym{\utb}{UTB}{University of Texas (Brownsville)}
\newacronym{\aps}{APS}{American Physical Society}
\newacronym{\lsu}{LSU}{Louisiana State University}
\newacronym{\rit}{RIT}{Rochester Institute of Technology}
\newacronym{\cfa}{CFA}{Center for Astrophysics}
\newacronym{\PCA}{PC}{Principal Component}
\newacronym{\ninja}{NINJA}{Numerical INJection Analysis}
\newacronym{\nrar}{NRAR}{Numerical Relativity Analytical Relativity}
\def\pn#1{post-Newtonian#1 (PN#1)\gdef\pn{PN}}
\def\qnm#1{quasi-normal mode#1 (QNM#1)\gdef\qnm{QNM}}
\def\isco#1{innermost stable circular orbit#1 (ISCO#1)\gdef\isco{ISCO}}
\def\eos#1{equation#1 of state (EOS#1)\gdef\eos{EOS}}
\def\tov#1{Tolman-Oppenheimer-Volkoff#1 (TOV#1)\gdef\tov{TOV}}
\def\ns#1{neutron star#1 (NS#1)\gdef\ns{NS}}
\def\WD#1{white dwarf#1 (WD#1)\gdef\WD{WD}}
\def\wdbh#1{white dwarf -- black hole#1 (WDBH#1)\gdef\wdbh{WDBH}}
\def\bbh#1{binary black holes#1 (BBH#1)\gdef\bbh{BBH}}
\def\bhns#1{black hole -- neutron star#1 (BHNS#1)\gdef\bhns{BHNS}}
\def\nsns#1{neutron star -- neutron star#1 (NSNS#1)\gdef\nsns{NSNS}}
\def\emri#1{Extreme Mass-Ratio Inspiral#1 (EMRI#1)\gdef\emri{EMRI}}
\def\emrb#1{Extreme Mass-Ratio Binaries#1 (EMRB#1)\gdef\emrb{EMRB}}
\def\sgrb#1{short gamma-ray burst#1 (SGRB#1)\gdef\sgrb{SGRB}}
\def\grb#1{gamma-ray burst#1 (GRB#1)\gdef\grb{GRB}}
\def\imbh#1{intermediate mass black hole#1 (IMBH#1)\gdef\imbh{IMBH}}
\def\smbh#1{supermassive black hole#1 (SMBH#1)\gdef\smbh{SMBH}}
\def\bh#1{black hole#1 (BH#1)\gdef\bh{BH}}
\def\ulx#1{ultra-luminous x-ray source#1 (ULX#1)\gdef\ulx{ULX}}
\def\lmxbs{low-mass x-ray Binaries (LMXBs)\gdef\lmxbs{LMXBs}\gdef\lmxb{LMXB}}
\def\lmxb{low-mass x-ray Binary (LMXB)\gdef\lmxbs{LMXBs}\gdef\lmxb{LMXB}}
\def\dns#1{double neutron star#1 (DNS#1)\gdef\dns{DNS}}
\def\mots#1{marginally outer trapped surface#1 (MOTS#1)\gdef\mots{MOTS}}

\newcommand\mathnew{\mathsurround=0pt}

\def\simov#1#2{\lower .5pt\vbox{\baselineskip0pt \lineskip-.5pt
        \ialign{$\mathnew#1\hfil##\hfil$\crcr#2\crcr\sim\crcr}}}
\def\MPR#1{{\it Moving Puncture Recipe}#1 (MPR#1)\gdef\MPR{MPR}}
\def\bh#1{black hole#1 (BH#1)\gdef\bh{BH}}
\def\ahz#1{apparent horizon#1 (AH#1)\gdef\ahz{AH}}
\def\bbh#1{binary black hole#1 (BBH#1)\gdef\bbh{BBH}}
\def\qnm#1{quasi-normal mode#1 (QNM#1)\gdef\qnm{QNM}}
\def\isco#1{innermost stable circular orbit#1 (ISCO#1)\gdef\isco{ISCO}}

\def\maya#1{\textsc{Maya}#1}
\def\et#1{\textsc{EinsteinToolkit}#1}

\def\maya#1{\textsc{Maya}#1}

\def\carpet#1{\textsc{Carpet}#1}

\def\ahf#1{\textsc{AHFinderDirect}#1}

\def\tde#1{tidal disruption event#1 (TDE#1)\gdef\tde{TDE}}
\def\whd#1{white dwarf#1 (WD#1)\gdef\whd{WD}}

\usepackage{xcolor}

\usepackage{iopams}
\begin{document}

\letter{Inside the Final Black Hole: Puncture and Trapped Surface Dynamics}
\author{Christopher Evans, Deborah Ferguson, Bhavesh Khamesra, Pablo Laguna, Deirdre Shoemaker}
\address{Center for Relativistic Astrophysics and School of
Physics, Georgia Institute of Technology,
Atlanta, GA 30332, USA}
\ead{cevans216@gatech.edu}

\vspace{10pt}
\begin{indented}
\item[]\today
\end{indented}

\begin{abstract}
A popular approach in numerical simulations of black hole binaries is to model black holes as punctures in the fabric of spacetime. The location and the properties of the black hole punctures are tracked with apparent horizons, namely outermost \mots{s}. As the holes approach each other, a common apparent horizon suddenly appears, engulfing the two black holes and signaling  the merger. The evolution of common apparent horizons and their connection with gravitational wave emission have been studied in detail with the framework of dynamical horizons. We present  a study of the dynamics of the \mots{s} and their punctures in the interior of the final black hole. The study focuses on head-on mergers for various initial separations and mass ratios. We find that \mots{s} intersect for most of the parameter space. We show that for those situations in which they do not, it is because of the singularity avoidance property of the moving puncture gauge condition used in the study. Although we are unable to carry out evolutions that last long enough to  show the ultimate fate of the punctures, 
our results suggest that \mots{s} always intersect and that at late times their overlap is only partial.  As a consequence, the punctures inside the \mots{s}, although close enough to each other to act effectively as a single puncture, do not merge. 
\end{abstract}

\section{\label{sec:introduction}Introduction}
An essential element in numerical relativity simulations involving \bh{s} is tracking the location and properties of the holes. A natural structure to accomplish this would be the event horizon of the \bh{}. The problem with these horizons is that they are teleological in nature; that is, we require knowledge of the entire space-time in order to identify their location and dynamics. The alternative is to track \ahz{s}~\cite{2004LRR.....7...10A} since to find them one only needs the intrinsic metric and extrinsic curvature of the spacetime hypersurface at a given time. \ahz{s} can be used to determine the mass and angular momentum of the \bh{s}~\cite{2003PRDDreyer}. Once the common \ahz{} forms during the merger, one can also estimate mass and spin multipole moments~\cite{2004CQGAshtekar,2013PRDAshtekar} to quantify the rate at which the final \bh{} approaches equilibrium and potentially identify when the ringdown phase begins~\cite{2018PRDGupta}. In addition, studies~\cite{2012PRDJaramilloa,2012PRDJaramillo,2012ACPJaramillo,2018PRDGupta,2020arXiv200306215P} have shown that fields at the \ahz{} are correlated with fields in the wave-zone and thus the \gw{} signal itself.

The fate of the common \ahz{} resulting from a \bh{} merger has been studied extensively and is fairly well understood~\cite{2003PRDAshtekar,2006PRDSchnetter,2013PRDAshtekar,2018PRDGupta}. Generally, after a common \mots{} forms it bifurcates into two surfaces. The outermost of these two surfaces expands and forms the \ahz{} for the final \bh{}, while the innermost surface contracts. Furthermore, the two \mots{s} that were initially the \ahz{s} of the two original \bh{s} continue to exist well after the formation of the common horizon. While these three interior \mots{s} have been studied in some detail~\cite{2006PRDSchnetter,2009ACPJaramillo,2015CQGMosta,2018PRDGupta,2019PRDPook-Kolba}, their ultimate fate remains uncertain and is the main focus of our work.

In this paper, we present a study to investigate the dynamics of  \mots{s} in the head-on collision of \bh{s} for various separations and mass ratios, with the holes modeled as punctures~\cite{PhysRevLett.78.3606}. We find that the \mots{s} of the merging punctures will in general intersect. For the situations in which they do not intersect, we show that it is due to the singularity avoidance properties of  the moving puncture gauge condition~\cite{2006PRLCampanelli,2006PRLBaker} used in the study. Our simulations are not long enough to show the ultimate fate of the \mots{s} and their punctures. At the same time, the results provide evidence that the punctures, although close enough to each other to act effectively as a single puncture, do not merge and the \mots{s} do not fully overlap. 
Section~\ref{sec:mots} provides a review of the \mots{} involved in the evolutions. In Section~\ref{sec:methods}, we outline the computational methods used for the  simulations. Results are given in Section~\ref{sec:results}, and Section~\ref{sec:conclusions} provides the conclusions.

\section{\label{sec:mots}Marginally Trapped Surfaces of Two Black Holes}
We consider a spacetime foliation of  spacelike  hypersurfaces $\Sigma_t$ labeled by a time parameter $t$.  The initial slice $\Sigma_0$ consists of two \bh{s} at a coordinate separation $d$. Depending on $d$, $\Sigma_t$ could have up to four \mots{s}~\cite{2006PRDSchnetter,2009ACPJaramillo,2015CQGMosta,2018PRDGupta,2019PRDPook-Kolba}. A \mots{} is a closed spacelike 2-surface in which the divergence of its outgoing null normals vanishes. An \ahz{} is the outer-most of the \mots{s}. For large enough $d$, $\Sigma_t$ will have two non-connected \mots{s}, $\mathcal{S}_{1}$ and $\mathcal{S}_{2}$, which correspond also to the \ahz{} of each individual \bh{}. At a separation $d_c$, a slice $\Sigma_t$ will also have a \mots{} $\mathcal{S}_{c}$  surrounding $\mathcal{S}_{1}$ and $\mathcal{S}_{2}$. Since $\mathcal{S}_{c}$ is now the outer-most \mots{}, the surface is an \ahz{}, called the \emph{common}   \ahz{}. In time, $\mathcal{S}_{c}$ will become the event horizon of the final \bh{.} For separations $d < d_c$, a  \mots{} $\mathcal{S}_{i}$ peels off from the interior of $\mathcal{S}_{c}$, shrinking and hugging $\mathcal{S}_{1}$ and $\mathcal{S}_{2}$. These four \mots{s},  $\mathcal{S}_{1}$, $\mathcal{S}_{2}$, $\mathcal{S}_{c}$, and  $\mathcal{S}_{i}$, are slices of four different dynamical horizons~\cite{2003PRDAshtekar,2004CQGAshtekar,2005ATMPAshtekar,2013PRDAshtekar}. However, in the case of $\mathcal{S}_{i}$, this only applies for a short time before the surface becomes timelike~\cite{2006PRDSchnetter}. For small initial separations, $\mathcal{S}_{1}$ and $\mathcal{S}_{2}$ are at all times nearly null surfaces~\cite{2006PRDSchnetter}, and thus  to good approximation they are isolated horizons~\cite{1999CQGAshtekar}.

A difficulty in studying the eventual fate of $\mathcal{S}_{1}$ and $\mathcal{S}_{2}$ is that, for the coordinate conditions typically used with punctures, the surfaces shrink after the formation of $\mathcal{S}_{c}$, requiring progressively finer spatial resolution to properly resolve them. As a result, it is challenging to make any definitive statements as to whether or not the \mots{s} exist based solely on the fact that they could not be located.

Many investigations into $\mathcal{S}_{1}$ and $\mathcal{S}_{2}$  have focused on locating them on a series of initial data slices varying their separations, avoiding the computational cost of performing high resolution simulations. Jaramillo, Ansorg, and Vasset~\cite{2009ACPJaramillo} studied Bowen-York~\cite{1980PRDBowen} initial data and found that at decreasing separations $\mathcal{S}_{1}$ and $\mathcal{S}_{2}$ merely shrink and show no indications of intersecting. Instead, $\mathcal{S}_{i}$ becomes highly distorted, and $\mathcal{S}_{1}$ and $\mathcal{S}_{2}$ `accumulate' against it. Pook-Kolb et al.~\cite{2019PRDPook-Kolba,2019PRLPook-Kolb,2019PRDPook-Kolb} studied an analogous series of time symmetric Brill-Lindquist initial data~\cite{1963PRBrill,1963JoMPLindquist} and found that $\mathcal{S}_{1}$ and $\mathcal{S}_{2}$  intersect and merge with $\mathcal{S}_{i}$ at the exact moment of intersection. Schnetter, Krishnan, and Beyer~\cite{2006PRDSchnetter} carried out simulations of head-on collision of Brill-Lindquist initial data. While they did lose the ability to track $\mathcal{S}_{i}$ rather early due to its high distortion, they made no statements about its ultimate fate or what happens to $\mathcal{S}_{1}$ and $\mathcal{S}_{2}$. They did however speculate that if these three surfaces do in fact merge, it is more likely that $\mathcal{S}_{1}$ and $\mathcal{S}_{2}$ merge first to form a new surface that then merges with $\mathcal{S}_{i}$.

\section{\label{sec:methods}Numerical Methods}
All simulations were done with our \maya{} code~\cite{2003VPPR5ICGoodale,2006CPCHusa,2012ApJHaas,2015ApJLEvans,2016PRDClark,2016CQGJani}, which is based on the BSSN formulation of the Einstein equations~\cite{1998PRDBaumgarte,2006PRLCampanelli}, with the moving punctures gauge condition~\cite{2006PRLCampanelli,2006PRLBaker} and  the \carpet{~\cite{2004CQGSchnetter,2006CQGSchnetter}} adaptive mesh refinement driver. The \maya{} code is our local version of the \et{} code~\cite{EinsteinToolkit:2019_10}. We use Brill-Lindquist initial data~\cite{1963PRBrill,1963JoMPLindquist} representing two  initially at rest, non-spinning \bh{s} with total mass $M = m_1 + m_2$, mass ratio $q = m_1/m_2$, and separated by a coordinate distance $d_0$. During the evolution, we use \ahf{~\cite{2004CQGThornburg}} to locate the \mots{s}.

As $\mathcal{S}_{1}$ and $\mathcal{S}_{2}$ approach each other, they will shrink in coordinate radius. The spatial resolution required to properly resolve and track them will thus increase accordingly. To ensure proper resolution, we activate additional refinement levels (one level each time the \mots{} radius reduces by half) to maintain roughly the same number of points within each \mots{}. When $\mathcal{S}_{c}$ is first located, $\mathcal{S}_{1}$ and $\mathcal{S}_{2}$ are each completely covered by three refinement levels, with each refinement having $60^3$ points. The resolutions for each refinement are: $M/100$, $M/200$ and $M/400$. Towards the end of a simulation, we activate up to five additional refinement levels, with a resolution of $M/12800$ at the finest level.

Our code solves the $\chi$ formulation of the BSSN equations~\cite{2006PRLCampanelli} and enforces a floor value $\chi \geq \chi_\epsilon$ to hande regions where the conformal factor diverges, e.g. at the punctures or singularities, where $\chi = 0$. We carried out a series of $q=1$ and fixed $d_0$  simulations in the range $10^{-3} > \chi_\epsilon >10^{-6}$. We found that the final time $t_f$ at which we cease locating $\mathcal{S}_{1}$ and $\mathcal{S}_{2}$ varied as $t_f \sim \chi_\epsilon^{-0.05}$. With a floor value of $\chi_\epsilon = 10^{-6}$, we obtain stable evolutions lasting $t_f \approx 20\,M$.

\begin{figure}
	\begin{center}
		\includegraphics[scale=0.5]{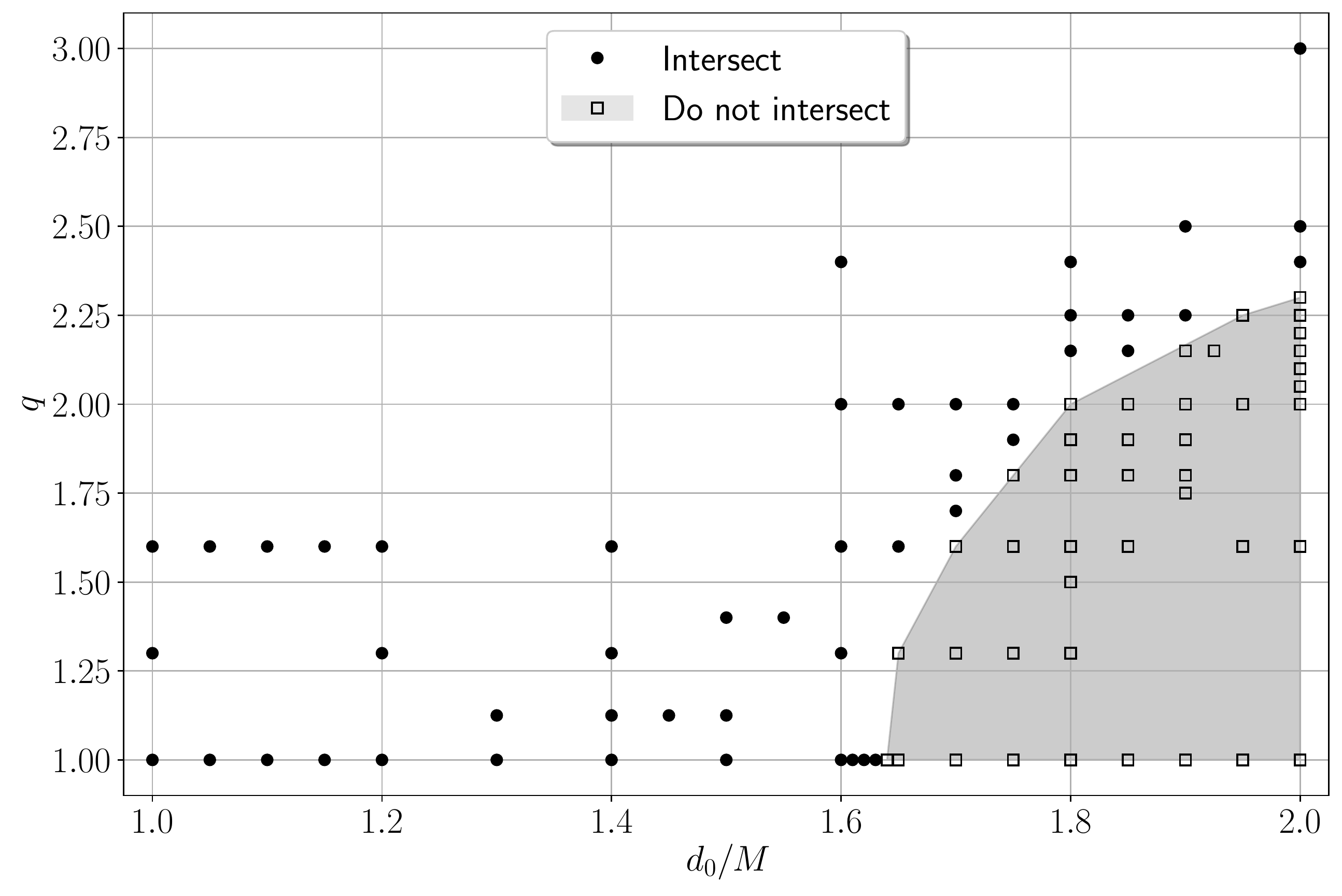}
		\caption{\label{fig:phase_diagram}Parameter space of simulations. Simulations are characterized by mass ratio $q$ and initial coordinate separation $d_0$. Cases for which $\mathcal{S}_{1}$ and $\mathcal{S}_{2}$ intersect are denoted by filled circles and non-intersecting by open boxes. In gray is the region of non-intersecting simulations.}
	\end{center}
\end{figure}

\section{\label{sec:results}Results}
Figure~\ref{fig:phase_diagram} shows the parameter space $q$ vs.\ $d_0$ of the simulations.   There are two distinct regions: one in the lower right corner (shaded gray)  in which $\mathcal{S}_{1}$ and $\mathcal{S}_{2}$ do not intersect and the rest in which they do. The  boundary separating these two regions is $q \approx 1.135 + \sqrt{4.065\,(d_0/M) - 6.674}$. 

\begin{figure}
	\begin{center}
		\includegraphics[scale=0.4]{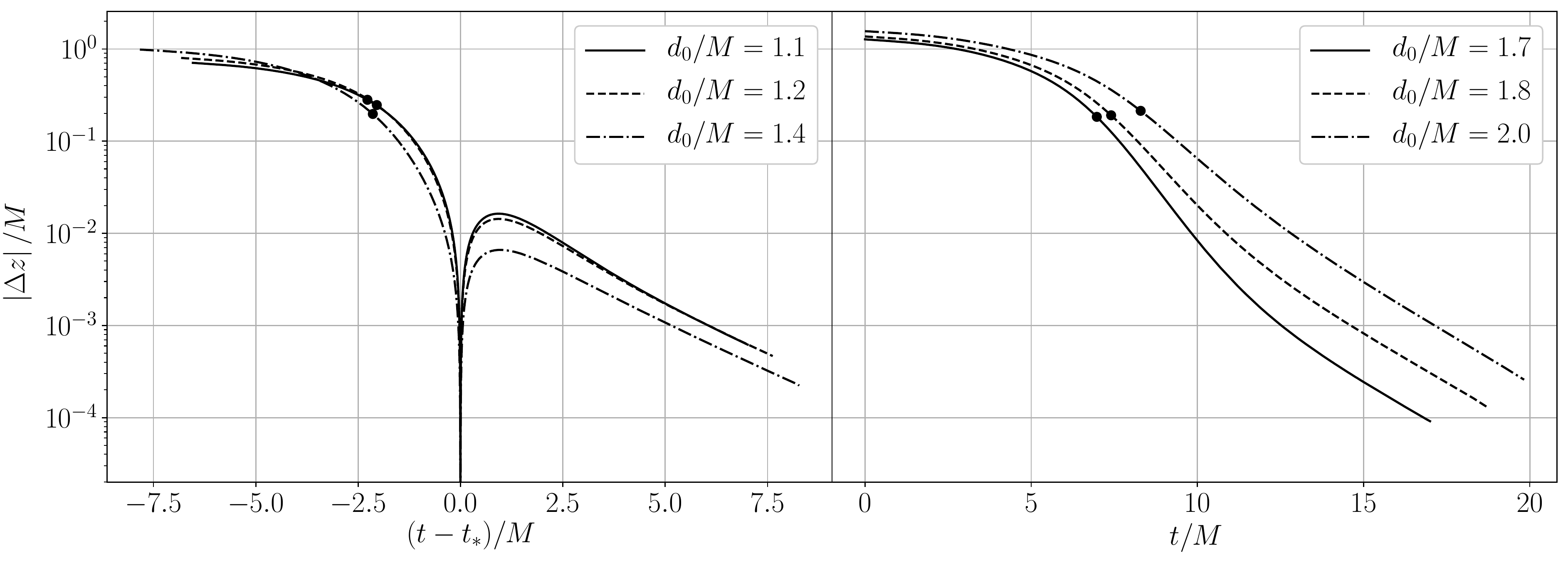}
		\caption{\label{fig:outcomes}Coordinate separation $\left| \Delta z \right|$ between $\mathcal{S}_1$ and $\mathcal{S}_2$ for a few $q=1$ examples as a function of coordinate time $t$. The left panel depicts three intersecting cases aligned at the time of intersection. The right panel shows three non-intersecting cases. Solid dots denote the time at which the common \ahz{} appears.}
	\end{center}
\end{figure}

Since all the simulations are head-on collisions with the holes along the $z$-axis, we track the coordinate separation between $\mathcal{S}_{1}$ and $\mathcal{S}_{2}$ with $\Delta z = z_1 - z_2$ where $z_1$ and $z_2$ are respectively the $z$-components at the surface of $\mathcal{S}_{1}$ and $\mathcal{S}_{2}$ that face each other. The coordinate origin is set at the center of mass of the initial configuration. Initially, $z_1>0$ and $z_2<0$. Thus, when the two surfaces intersect, $z_1 < z_2$, and $\Delta z$ becomes negative. Figure~\ref{fig:outcomes} shows for a few $q=1$ cases $\left| \Delta z \right|$ as a function of coordinate time $t$. The left panel shows three cases in which $\mathcal{S}_1$ and $\mathcal{S}_2$ intersect. The time axis has been shifted so the cases align when the surfaces intersect at time $t_*$, which depends on $d_0$. The right panel shows three other cases in which $\mathcal{S}_1$ and $\mathcal{S}_2$ do not intersect. The panels show that at late times the separation for the non-intersecting cases and the overlap for the intersecting cases both decrease as $\left| \Delta z \right| \sim e^{- t/\lambda}$. The same exponential decay extends to the $q \ne 1$ cases. For all cases, we find that  $\lambda \approx 2M$. The exponential decay in the surface separation is also present in the coordinate separation, $d$, of the two punctures. Shortly after the formation of $\mathcal{S}_{c}$, we find again that $d \propto e^{-t/\lambda}$  with $\lambda \approx 2M$. 

To understand the exponential decay in the separation between $\mathcal{S}_1$ and $\mathcal{S}_2$ as well as between the punctures, we recall that in the moving puncture gauge the lapse function $\alpha$ satisfies the $1+\log$ type slicing condition:
$
  \left( \partial_t - \beta^i \partial_i \right) \alpha = -n \alpha K,
$
with $\beta^i$ the shift vector, $K$ the trace of the extrinsic curvature, and $n$ a constant. As is customary, we choose $n = 2$. With this choice, stationary slices of a single Schwarzchild puncture are given by a family of \emph{trumpet slices}~\cite{2007PRLHannam}, for which the surface of zero isotropic radius (the trumpet surface) has a non-zero areal radius, and the lapse on the trumpet surface vanishes, thus avoiding the singularity at the puncture. With the moving puncture gauge,
the position of the punctures $x^i_{1,2}$ are found from integrating  $\partial_t\, x^i_{1,2} = -\beta^i_{1,2}$~\cite{2006PRLCampanelli}. Since for the Schwarzchild trumpet slices, $\beta^r = r/\lambda$ near the puncture~\cite{2007PRLHannam,2009GRGBrugmann}, the radial coordinate distance to each puncture is given by $r_{1,2} \propto e^{-t/\lambda}$, with the decay rate $\lambda$ computed from
$
  \lambda^2 = \frac{1}{r} \beta^r \partial_r \beta^r.
$
Substituting the solution for $1+\log$ trumpet slices found by Hannam et al.~\cite{2008PRDHannama} into this expression yields
$
 ( \lambda/M)^2 \approx (R_0/M)^3/(2 - R_0/M) \,,
$
in which $R_0 \approx 1.3124\,M$. This gives $\lambda \approx 1.82\;M$, which is consistent with our numerical value from our simulations. The minor disagreement is easily explained by the fact that our numerical simulations do not reach full stationarity before completion and the shift vector is evaluated slightly away from the puncture.

\begin{figure}
	\begin{center}
		\includegraphics[scale=0.4]{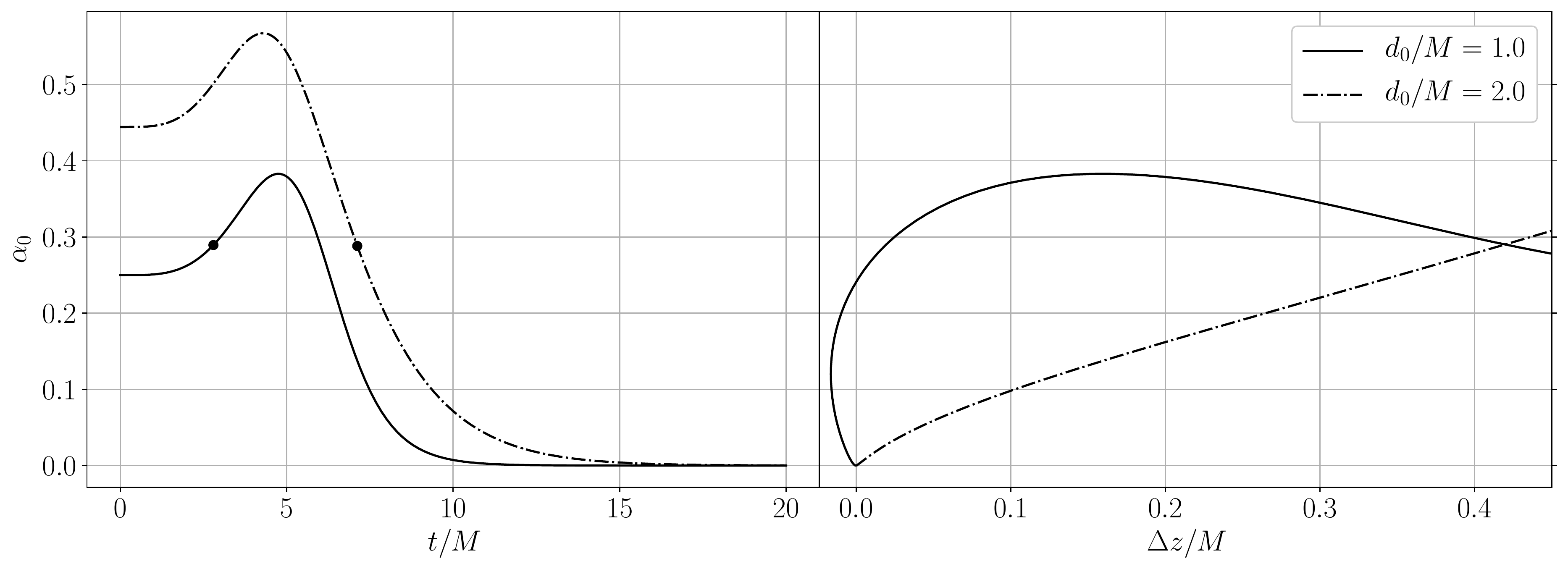}
		\caption{\label{fig:lapse_collapse}Lapse at the coordinate origin $\alpha_0$ for $q=1$ and $d_0/M = 1,\, 2$. The left panel shows how $\alpha_0$ changes as a function of coordinate time $t$. The right panel shows how  $\alpha_0$ changes with $\Delta z$. For $d_0/M = 2$, by the time $\Delta z = 0$,  the lapse has already collapsed. In contrast, for $d_0/M = 1$, $\Delta z = 0$ is reached when $\alpha_0 \approx 0.25$. After this point, as $\alpha_0$ collapses, $\Delta z$ reaches a minimum and at late times $\Delta z \rightarrow 0$.}
	\end{center}
\end{figure}

To demonstrate how the lapse function is connected to the behavior observed in  $\mathcal{S}_1$ and $\mathcal{S}_2$, we show in Figure~\ref{fig:lapse_collapse} the lapse function at the origin, $\alpha_0$, as a function of coordinate time $t$ (left panel) and as function of $\Delta z$ (right panel). Two cases are plotted: one in which the surfaces do not intersect ($d_0/M = 2$) and another in which they do ($d_0/M = 1$). It is clear from the left panel that in both cases the lapse eventually collapses and thus halts the evolution. The difference on how the collapse proceeds in each case and affects the final outcome is more evident in the right panel. We see in this panel that at $\Delta z \approx 0.42$ both cases are at the same separation. The solid dots in the left panel label $\alpha_0$ at this separation. For the non-intersecting ($d_0/M = 2$) case, $\alpha_0$ is already starting to collapse. On the other hand, for the intersecting  ($d_0/M = 1$) case, $\alpha_0$ is still growing; thus, the evolution lives longer and the surfaces are able to intersect before the end of the simulation. What is also interesting is that the degree of intersection or surface overlap reaches a maximum and then decreases as the lapse enters collapse. As we will show later, this is a coordinate effect.

\begin{figure}
	\begin{center}
		\includegraphics[scale=0.4]{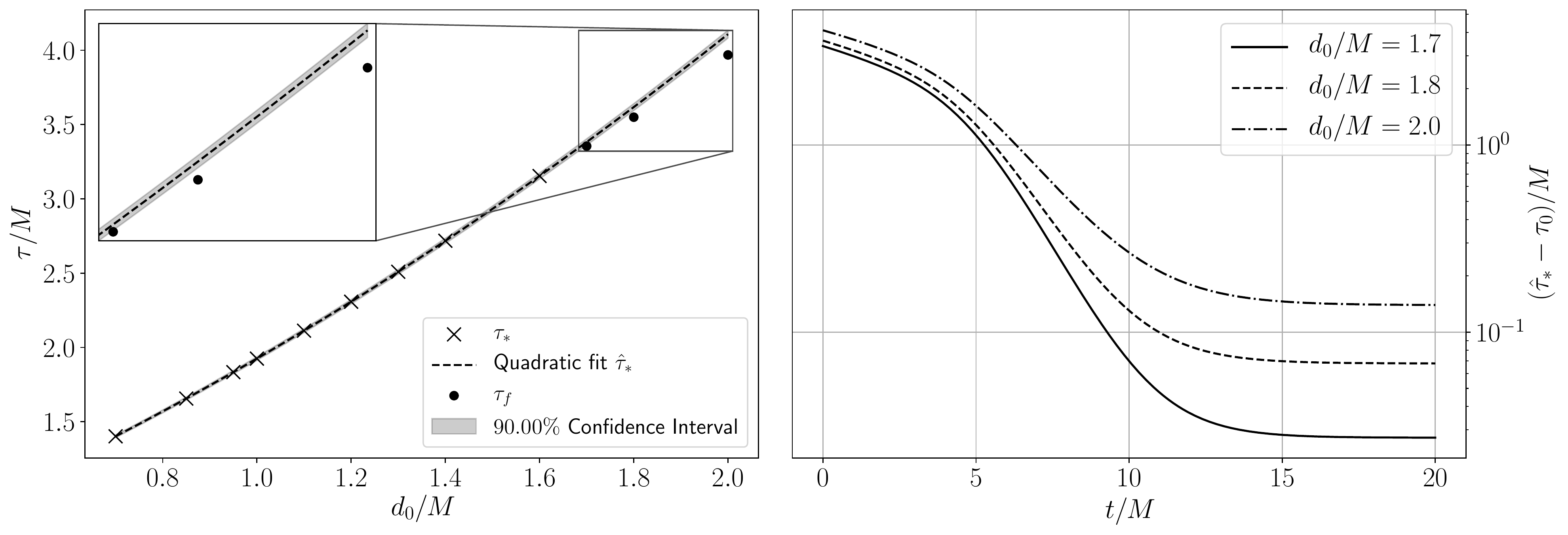}
		\caption{\label{fig:proper_time_cross} Left panel shows proper time $\tau_*$  (crosses) at the origin when $\mathcal{S}_1$ and $\mathcal{S}_2$ intersect at time $t_{*}$ for $q=1$ as a function of $d_0$. Included is also a quadratic fit $\hat{\tau}_{*}$ and in gray the $90\%$ confidence interval. With solid dots are the proper time $\tau_f$ elapsed at the origin by the end of the simulations  for three non-intersecting cases. The right panel shows $\hat \tau_* - \tau_0$ as a function of $t$ for those three non-intersecting cases. }
	\end{center}
\end{figure}

To further support the view about the effect of the lapse, we have measured the proper time
$ \tau_0(t) = \int^t_{t=0} \alpha_0\,d t'$ at the origin. The left panel in Fig.~\ref{fig:proper_time_cross} shows with crosses $\tau_* \equiv \tau_0(t_{*})$, where $t_*$ is the time when $\mathcal{S}_1$ and $\mathcal{S}_2$ intersect. Also plotted is a quadratic fit $\hat{\tau}_{*}/M = 0.337\,(d_0/M)^2 +1.170\,(d_0/M) + 0.417$ and in gray the $90\%$ confidence interval. The insert shows extrapolation of  $\hat{\tau}_{*}/M$ beyond the intersecting cases, with three data points (solid dots) denoting non-intersecting cases in which $\tau_f = \tau_0(t_f)$, with $t_f$  the time at the end of the simulation. Notice that $\tau_f < \hat\tau_*$, suggesting that, if in those cases the evolution had lasted $\hat \tau_* - \tau_f$  longer, the surfaces would have intersected. The right panel in Fig.~\ref{fig:proper_time_cross} shows $\hat \tau_* - \tau_0$ as a function of coordinate time $t$ for the three non-intersecting cases. Notice that $\hat \tau_* - \tau_0 \rightarrow$ constant, as the lapse collapses, signaling that the progression of proper time has halted.

\begin{figure}
	\begin{center}
		\includegraphics[scale=0.4]{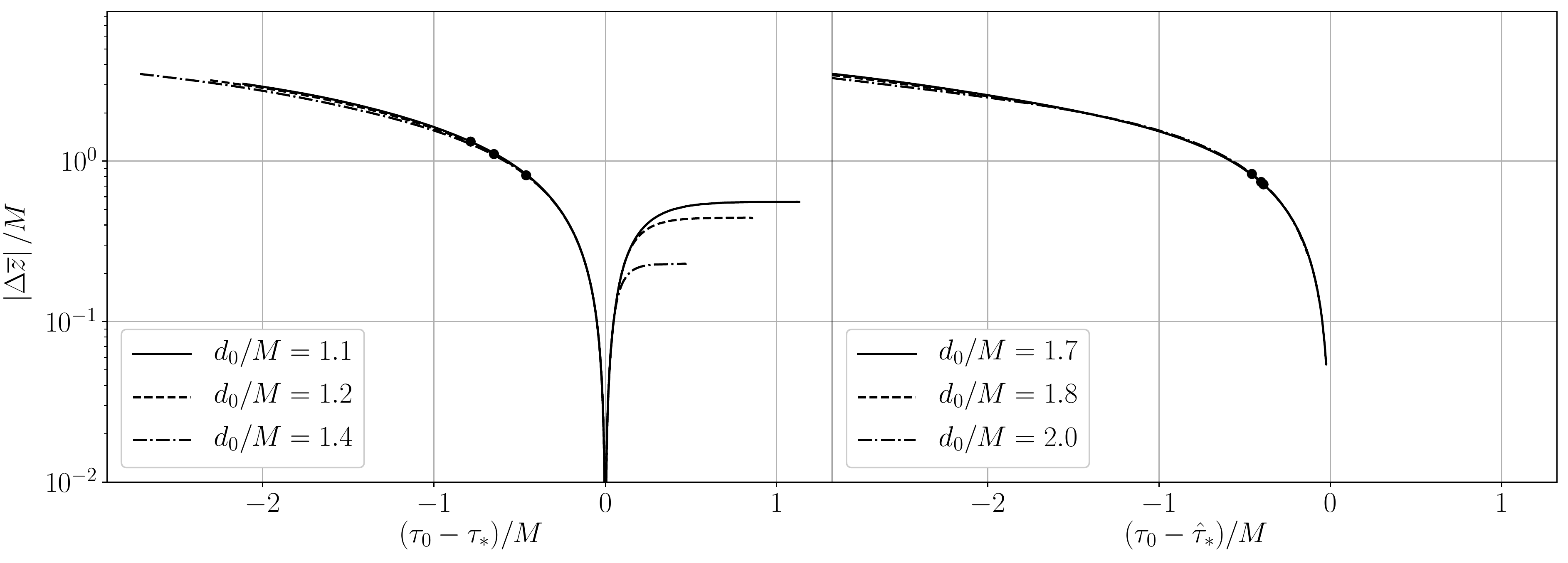}
		\caption{\label{fig:outcomes2}Proper separation $\left| \Delta \bar z \right|$ between $\mathcal{S}_1$ and $\mathcal{S}_2$ for a few $q=1$ examples as a function of proper time $\tau_0$ measured at the origin of the coordinate system. The left panel depicts three intersecting cases aligned at the time of intersection. The right panel shows three non-intersecting cases. Solid dots denote the time at which the common \ahz{} appears.}
	\end{center}
\end{figure}

As mentioned before, tracking the separation or overlap of $\mathcal{S}_1$ and $\mathcal{S}_2$ with coordinate distances has the complication that the choice of gauge influences the outcome. To circumvent this, we show in Figure~\ref{fig:outcomes2} the proper distance separation $| \Delta \bar{z} |$  as a function of proper time $\tau_0$ corresponding to the cases in Fig.~\ref{fig:outcomes}. The left panel depicts the intersecting cases with time shifted by the time at intersection, $\tau_*$. The right panel shows three non-intersecting cases with the time also shifted but in this case by $\hat\tau_*$ from the fit in Fig.~\ref{fig:proper_time_cross}. It is clear from both panels that for  $\tau_0-\tau_* < 0$ and $\tau_0-\hat\tau_* < 0$, the proper separation is independent of $d_0$. Also, if we were to combine the data from both panels, it would show that for these times all cases lie on top of each other; thus, there is no difference between intersecting and non-intersecting cases. Therefore, here again the data suggest that, if the evolutions for the non-intersecting cases had proceeded, the surfaces would have eventually intersected. The differences in $| \Delta \bar{z} |$  with $d_0$ arise when $\tau_0 - \tau_*>0$, namely when the surfaces overlap. The left panel shows that at late times $\mathcal{S}_1$ and $\mathcal{S}_2$ reach a constant proper overlap, the smaller the value of $d_0$ the larger the overlap. Furthermore, we find that the final overlap volume is never large enough to contain the punctures; they remain in the non-overlapping regions. 

\begin{figure}
  \begin{center}
    \includegraphics[scale=0.4]{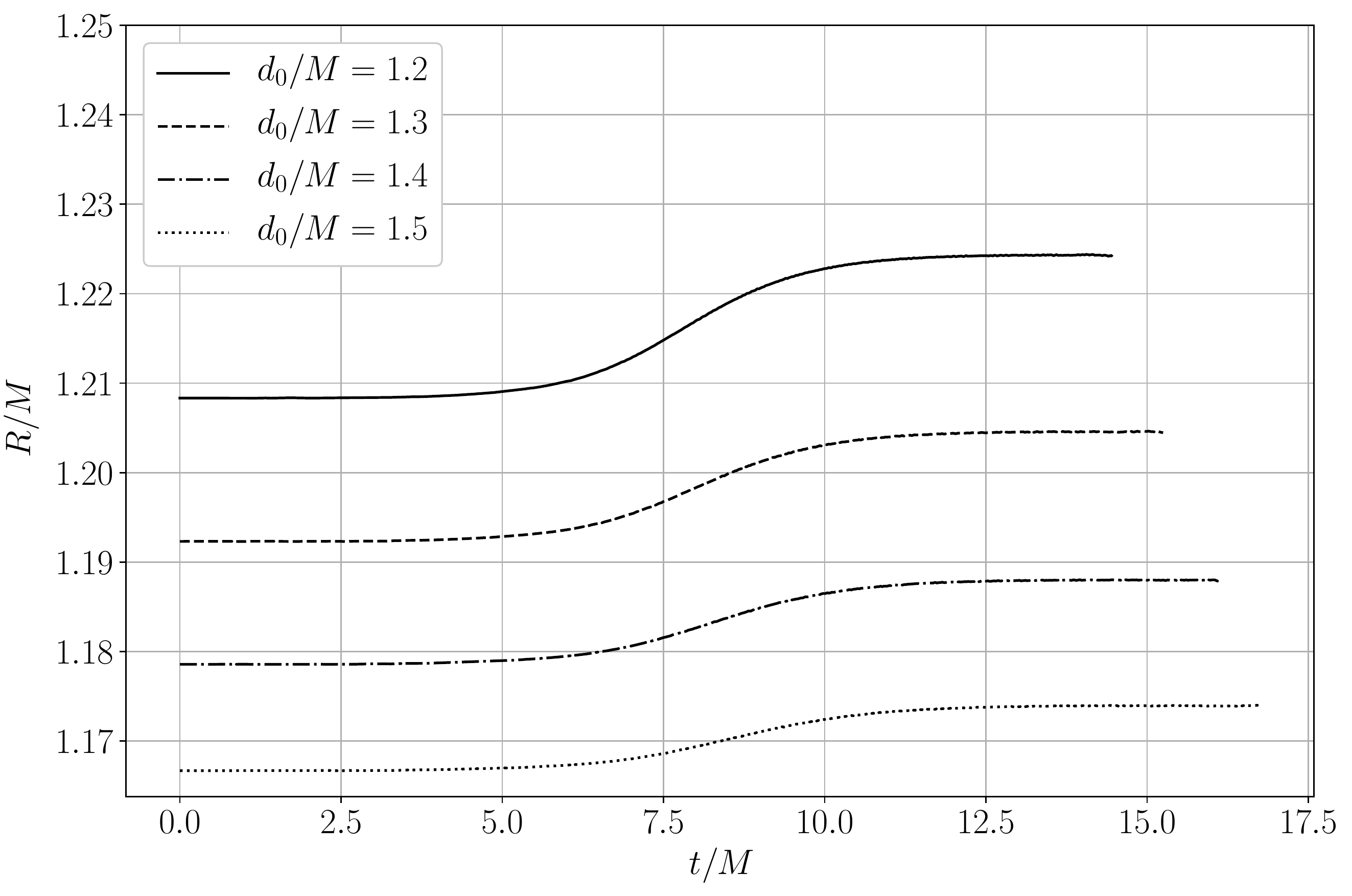}
    \caption{\label{fig:area_plot} Evolution of the areal radius of  $\mathcal{S}_1$ for  intersecting cases with $q=1$. The areal radius of $\mathcal{S}_2$ is the same as that for $\mathcal{S}_1$ since the holes have equal masses.}
  \end{center}
\end{figure}

To gain further insight about the final state of the \mots{s} and punctures, we show in Fig.~\ref{fig:area_plot} the evolution of the areal radius $R = \sqrt{A/4\pi}$ with $A$ the area of the \mots{} for a few intersecting cases with $q=1$. It is clear that toward the end of the simulation, the surfaces $\mathcal{S}_1$ and $\mathcal{S}_2$ reach a constant areal radius and thus become isolated horizons~\cite{2004LRR.....7...10A}. This together with the finding that $\mathcal{S}_1$ and $\mathcal{S}_2$ have a constant proper overlap strongly suggest that the configuration is essentially frozen and the punctures will not merge.

\section{\label{sec:conclusions}Conclusions}

We have presented results from a two-parameter study (mass ratio $q$ and initial separation $d_0$) of head-on collisions of \bh{s}. The focus was on the ultimate fate of the \mots{s}~$\mathcal{S}_1$ and $\mathcal{S}_2$ that initially were the \ahz{s} of the colliding \bh{s}.  Depending on the values of $q$ and $d_0$, once inside the common \ahz{} the surfaces $\mathcal{S}_1$ and $\mathcal{S}_2$ intersect if the lapse function $\alpha_0$ takes longer to collapse before the end of the simulation. The collapse of the lapse is intrinsic to the singularity avoidance properties of the moving puncture gauge condition~\cite{2006PRLCampanelli,2006PRLBaker} used in the simulations. We find that at late times for all cases, the coordinate separation of the \bh{} punctures and of the \mots{} surfaces $\mathcal{S}_1$ and $\mathcal{S}_2$ decrease $ \propto e^{-t/\lambda}$  with $\lambda \approx 2M$. When the separation of $\mathcal{S}_1$ and $\mathcal{S}_2$ is measured by proper distances, we find that at early times all cases exhibit the same behavior as a function of proper time. The data suggest that, if it were not for the collapse of the lapse, all cases would intersect. Furthermore, at late times the intersection or overlap freezes. Similarly, at late times, the areal radius of $\mathcal{S}_1$ and $\mathcal{S}_2$ reach a constant, thus becoming isolated horizons. These two facts, the freezing of the areal radius and the overlap, strongly suggest that the punctures do not merge. However, since this occurs at very small separations, ($|\Delta z| \sim 10^{-4}M$), for practical purposes, the two punctures act as a single puncture, namely the singularity of the final \bh{.} 

\section*{Acknowledgments} We would like to thank Abhay Ashtekar for useful discussions. This work was supported by NSF grants 1806580, 1550461, and XSEDE  allocation TG-PHY120016.

\section*{References}
\bibliographystyle{unsrt}
\bibliography{references}

\end{document}